\documentclass[a4paper,11pt]{article}
\usepackage{pos}

\renewcommand{\hookAfterAbstract}{%
\par\bigskip\bigskip
\textsc{CERN-TH-2023-185}
}

\newcommand{\eps}{\ensuremath \epsilon}
\renewcommand{\i}{\ensuremath{\mathrm{i}}}

\newcommand{\Li}{\ensuremath{\mathrm{Li}}}

\usepackage{cleveref}
\usepackage{comment}

\title{Two-loop five-particle scattering amplitudes}

\author*[a]{Simone Zoia}

\affiliation[a]{CERN, Theoretical Physics Department, CH-1211 Geneva 23, Switzerland}

\emailAdd{simone.zoia@cern.ch}

\abstract{I discuss the recent advances in the computation of two-loop scattering amplitudes for five-particle processes. The latter are fundamental ingredients to obtain predictions at the next-to-next-to-leading order (NNLO) in QCD for many interesting LHC processes. I discuss the state-of-the-art technology for computing scattering amplitudes analytically, and present new results relevant for the LHC phenomenology.}

\FullConference{
  16th International Symposium on Radiative Corrections: Applications of Quantum Field Theory to Phenomenology (RADCOR2023)\\
  28th May -- 2nd June, 2023\\
  Crieff, Scotland, UK
}

\begin{document}
\maketitle

\section{Introduction}
\label{sec:Introduction}

Improvements in systematic uncertainties and the increasingly large amount of data collected at the Large Hadron Collider (LHC) mean that measurements at the percent-level precision will soon be available for a wide class of observables. Exploiting fully this wealth of data requires that we keep the theoretical uncertainties in line with the experimental ones. This is a multi-faceted challenge which requires improvements on all levels of theoretical predictions. One crucial step to this end is the computation of fixed-order predictions at least at the Next-to-Next-to-Leading Order (NNLO) in QCD.
This is particularly challenging for processes involving many physical scales, which are however of great interest since the high energy of the interactions at the LHC can lead to the production of many particles.

The current frontier for multi-particle NNLO predictions are $2\to3$ processes. The main bottleneck is the computation of the two-loop five-particle scattering amplitudes required for the double-virtual corrections.
Over the last years we have seen important progress in this direction, building upon the campaign which mastered the $2\to2$ processes and a deepened understanding of the polylogarithmic special functions appearing in the amplitudes. 
In the case of fully massless final states, all partonic processes needed for three-jet (in the leading colour approximation)~\cite{Abreu:2019odu, Abreu:2021oya}, three-photon~\cite{Chawdhry:2019bji, Abreu:2020cwb, Chawdhry:2020for, Abreu:2023bdp}, di-photon + jet~\cite{Agarwal:2021grm, Chawdhry:2021mkw}, and photon + di-jet~\cite{Badger:2023mgf} production at the LHC have been computed. 
More recently, several leading-colour amplitudes involving an external massive vector boson~\cite{Hartanto:2019uvl, Abreu:2021asb, Badger:2021nhg, Badger:2021ega, Badger:2022ncb, Hartanto:2022qhh} have also become available, and the first steps have been taken for processes with internal massive propagators~\cite{Badger:2022mrb,Badger:2022hno} (see Matteo Becchetti's talk~\cite{talk-becchetti}). 
Remarkably, for most of these $2\to3$ processes, theoretical predictions at NNLO in QCD followed within months from the computation of the two-loop amplitudes~\cite{Chawdhry:2019bji, Kallweit:2020gcp, Chawdhry:2021hkp, Czakon:2021mjy, Badger:2021ohm, Chen:2022ktf, Hartanto:2022qhh, Hartanto:2022ypo, Buonocore:2022pqq, Badger:2023mgf, Alvarez:2023fhi}. 
On the one hand, this shows that the availability of the two-loop amplitudes is indeed the bottleneck for such computations. On the other hand, this proves that the analytic expressions obtained for them meet the high demands of phenomenology in terms of stability and efficiency of the numerical evaluation. 

In order to understand what kind of difficulties need to be overcome, it is instructive to have a look at how the analytic expression of a loop amplitude looks like. 
It is a function of the particles' momenta,\footnote{For simplicity we here neglect the dependence on the space-time dimension $D$.} which we denote by $p$, 
with the following general structure,
\begin{align}
\mathcal{A}(p) = \sum_i c_i(p) \times f_i(p) \,,
\end{align}
where $f_i(p)$ are special functions, i.e.\ functions with branch cuts such as logarithms and polylogarithms, and $c_i(p)$ are rational coefficients. This separation allows us to distinguish two types of complexity: algebraic and analytic. The \emph{algebraic complexity} affects the rational coefficients: while they are simple from the analytic point of view, their expressions can be exceedingly large. To give an idea, in the massless five-particle processes they are functions of five independent variables and, written as ratios of polynomials, their maximum polynomial degree can exceed 100. Manipulating these coefficients symbolically is thus difficult from the computational point of view. For the special functions the difficulty is not only computational, but also conceptual. Understanding what class of special functions is relevant for a given amplitude is often challenging, and evaluating them even at a single phase-space point may be problematic due to their intricate branch-cut structure. Developing the mathematical technology to bring under control this \emph{analytic complexity} is thus imperative in order to obtain analytic results that can be used for phenomenology.

For the fully massless five-particle processes and for the five-particle processes with a single external massive particle, the problem of the analytic complexity is completely solved at two-loop order. The solution consists in representing the appearing Feynman integrals in terms of bases of algebraically independent special functions, called \emph{pentagon functions}~\cite{Gehrmann:2018yef,Chicherin:2018old,Chicherin:2020oor,Chicherin:2021dyp,Abreu:2023rco}, which allow for a compact analytic form as well as a stable and efficient numerical evaluation. Dmitry Chicherin discusses thoroughly this topic in his talk~\cite{talk-chicherin}. 

In my talk, I focus on the techniques we developed to tame the algebraic complexity, and use as concrete example the two-loop amplitudes required for the NNLO QCD corrections to isolated photon production in association with two jets in hadron collisions~\cite{Badger:2023mgf}.

\section{Amplitude workflow}
\label{sec:Workflow}

In this section I present a workflow for computing scattering amplitudes analytically based on finite-field arithmetic~\cite{vonManteuffel:2014ixa,Peraro:2016wsq}. 
It was developed in a long series of works by a multitude of authors including myself, and has proven efficient throughout a wide range of cutting-edge applications~\cite{Peraro:2016wsq,Badger:2017jhb,Badger:2018enw,Badger:2019djh,Peraro:2019svx,Hartanto:2019uvl,Badger:2021owl,Badger:2021nhg,Badger:2021ega,Badger:2022mrb,Badger:2022ncb,Hartanto:2022qhh,Badger:2023xtl,Badger:2023mgf}. I will give here a brief and general overview of the method, considering the computation of a partial helicity amplitude in general.\footnote{This methodology can be similarly applied to the computation of any quantity involving Feynman integrals.} I refer the readers who are not familiar with the modern on-shell methods for scattering amplitudes to refs.~\cite{Mangano:1990by,Dixon:1996wi,Badger:2023eqz}.

The starting point is the expression of the chosen helicity amplitude $A$ in terms of Feynman diagrams, which we generate using \texttt{QGRAF}~\cite{Nogueira:1991ex}. It is a function of a certain set of independent kinematic variables, which we label cumulatively by $\vec{x}$, and of the dimensional regulator $\eps = (4-D)/2$, where $D$ is the number of space-time dimensions. We use a momentum-twistor parametrisation for the kinematics~\cite{Hodges:2009hk,Badger:2016uuq}, which allows us to represent also the helicity spinors and rationalises certain square roots that appear in the kinematics of more than four particles. We process the Feynman-diagram expression to extract the chosen partial amplitude, and rewrite it as a linear combination of rational coefficients $c_i$ and scalar Feynman integrals $I_i$,
\begin{align} \label{eq:A2I}
A\left(\vec{x} , \epsilon\right) = \sum_i c_i(\vec{x}, \epsilon) \, I_i(\vec{x}, \epsilon) \,.
\end{align}

The scalar Feynman integrals satisfy linear relations, for instance the integration-by-parts identities (IBPs)~\cite{Tkachov:1981wb,Chetyrkin:1981qh}. 
Only a subset of them, called master integrals (MIs), are linearly independent. The next step is then to rewrite all integrals in terms of MIs, which I denote by $G_j$:
\begin{align} \label{eq:I2G}
I_i\left(\vec{x} , \epsilon\right) = \sum_j W_{ij}(\vec{x}, \epsilon) \, G_{j}(\vec{x}, \epsilon) \,.
\end{align}
This task is in principle straightforward using Laporta's algorithm~\cite{Laporta:2001dd}. It amounts to the generation and solution of a large set of linear relations among the scalar integrals. Using \cref{eq:I2G} we can then express the amplitude in terms of MIs:
\begin{align} \label{eq:A2G}
A\left(\vec{x} , \epsilon\right) = \sum_i d_i(\vec{x}, \epsilon) \, G_i(\vec{x}, \epsilon) \,.
\end{align}
In practice, however, these steps suffer from a common plague of symbolic computations: intermediate expression swell. While the input (the Feynman diagrams) and the final result for the amplitude are comparatively compact ---~in particular the latter, as it is constrained by many physical properties~--- the intermediate expressions grow to the point of hindering the computation.

Two simple but crucial insights allow us to overcome this issue. First, we do not need to know analytically the intermediate expressions, such as the coefficients $W_{ij}$ in \cref{eq:I2G}; we only care about the final result, the amplitude. Second, the rational coefficients collapse to numbers once they are evaluated numerically, deflating the expression swell. Leveraging these two points, ref.~\cite{Peraro:2016wsq} pioneered an approach which trades the symbolic manipulation of the rational functions for numerical evaluations over finite fields ---~that is, using integers modulo a (large) prime number~--- throughout the entire computation. Using finite fields allows us to avoid the loss of accuracy inherent to floating-point numbers, as well as the computationally expensive arbitrary-precision arithmetic required to work with exact rational numbers. The analytic expression can then be recovered from sufficiently many numerical samples using multivariate functional reconstruction algorithms.

Our entire workflow therefore becomes a ``black box'' which, for given values of the kinematic variables $\vec{x}$ and of $\eps$, returns the values of the coefficients of the MIs $d_i$ in \cref{eq:A2G}, performing all computations numerically over finite fields. To this end, we use the multi-purpose \texttt{Mathematica}/\texttt{C++} framework \texttt{FiniteFlow}~\cite{Peraro:2019svx}. 
In principle we could now reconstruct the analytic expression of the MI coefficients $d_i$ by evaluating the black box sufficiently many times. The number of numerical samples depends on the complexity of the rational functions which need to be reconstructed. In order to make the most of the finite-field approach, therefore, one should reconstruct something that is as simple as possible. To this end, we also Laurent-expand around $\eps = 0$ the coefficients $d_i(\vec{x},\eps)$ and the MIs $G_i(\vec{x},\eps)$, truncating the expansion at the required order. While the expansion of the coefficients can be performed within the finite-field setup, expanding the MIs requires having analytic expressions in terms of special functions for them. 

The analytic computation of the MIs is typically the most difficult step. As we are here focusing on the problem of the algebraic complexity, I will assume that analytic expressions for the MIs are available, and refer to Dmitry Chicherin's talk~\cite{talk-chicherin} for a discussion of this issue. For our purposes, it suffices to mention two important aspects. First, in order to completely separate the algebraic from the analytic complexity, and the rational coefficients from the special functions in the amplitudes, the MIs must be chosen so that they satisfy differential equations in the canonical form~\cite{Henn:2013pwa} (see refs.~\cite{Henn:2014qga,Badger:2023eqz} for an introduction to this concept). In practice, this means that the MIs are given ---~order by order in $\eps$~--- by polynomials in the special functions (and transcendental constants) with constant rational coefficients. A toy example for a Feynman integral of this form with a single kinematic variable $z$ would look like
\begin{align}
G_j(z, \eps) \approx 1 + \eps \, \left( \log(z) - \i \pi \right) + \eps^2 \, \left( \Li_2(z) + \log^2(z) + \frac{\pi^2}{6}\right) + \mathcal{O}\left(\eps^3\right)\,,
\end{align}
where $\Li_2$ is the dilogarithm. The second important aspect for our purposes is that special functions satisfy functional identities. A simple example is given by the dilogarithm's inversion formula,
\begin{align}
\Li_2(z) + \frac{1}{2} \log^2(-z) + \Li_2\left(\frac{1}{z}\right) + \frac{\pi^2}{6} = 0  \,,
\end{align}
for $z<0$. Unlike the rational coefficients, the special functions are treated symbolically, and the identities among them must be known exactly in order to reconstruct a final result that is as simple as possible. This is why it is important that we express the MIs in terms of a basis\footnote{Strictly speaking, we should talk of a generating set rather than a basis.} of \emph{algebraically independent} special functions. This gives a unique representation for all MIs and the amplitude where all cancellations and simplifications due to the special functions are built in. 

Denoting the list of special functions which form a basis by $\vec{f}(\vec{x})$, our final expression for the helicity partial amplitude has the form
\begin{align} \label{eq:A2f}
A\left(\vec{x} , \epsilon\right) = \sum_{w = - 2 L}^{w_{\rm max}} \eps^w \sum_i e_i^{(w)}(\vec{x}) \ \mathrm{mon}_i \left[ \vec{f}(\vec{x})\right] \,,
\end{align}
where $w_{\mathrm{max}}$ depends on the intended application of our results, and $\mathrm{mon}_i \left[ \vec{f}(\vec{x})\right]$ are monomials in the special functions. We may now evaluate the rational coefficients $e_i^{(w)}(\vec{x})$ over finite fields, and reconstruct their analytic expression using \texttt{FiniteFlow}'s functional reconstruction algorithms. While evaluating the coefficients at a single phase-space point $\vec{x}$ is typically feasible, evaluating them sufficiently many times for the reconstruction to succeed can still be prohibitive. In  \cref{sec:Reconstruction} I will discuss a number of strategies which allow us to leverage our physical understanding of the analytic structure of the amplitudes to optimise substantially the reconstruction.

\section{Functional reconstruction strategy}
\label{sec:Reconstruction}

In this section I review a number of techniques to optimise the functional reconstruction of a set of rational coefficients $\{r_i(\vec{x})\}$ from their numerical values over finite fields. A proxy for the difficulty of this problem is given by the total time it takes to collect the samples required for the reconstruction.\footnote{The time required to perform the actual reconstruction from the numerical samples is negligible.} This can be estimated as
\begin{align}
\text{total evaluation time} \approx \frac{\text{evaluation time of one point} \ \times \ \# \ \text{points}}{\# \ \text{CPUs}} \,.
\end{align}
Each of the factors on the right-hand side offers a different line of attack.

\smallskip 

First of all, the evaluations can be parallelised, which is why the total reconstruction time is inversely proportional to the number of available CPUs.
The availability of computing resources is thus a crucial factor in this game.

\smallskip 

The evaluation time of the coefficients at one random point in the finite field is mainly affected by the evaluation of the solution to the IBPs.
The traditional way to generate the IBPs introduces integrals which have higher propagator powers with respect to those appearing in the amplitudes. These ``spurious'' integrals increase considerably both the size and the number of variables of the linear system of equations we need to solve to perform the reduction of the amplitudes. This makes the evaluation of the solution over finite fields slower, and increases the memory load, this way reducing the number of CPUs which can be used in parallel on a single machine.
To overcome these issues, we generated relations among integrals without higher propagator powers~\cite{Gluza:2010ws,Ita:2015tya,Larsen:2015ped}. 
We made use of the Baikov parameterisation of Feynman integrals~\cite{Baikov:1996rk,Baikov:1996cd}, where IBP relations typically contain both higher propagator powers and dimensionally-shifted integrals. 
We derived equations without dimensionally-shifted integrals as closed-form solutions to polynomial equations called syzygy equations~\cite{Bohm:2017qme}. Next, we put these solutions into a sparse matrix, and eliminated the higher propagator powers through Gaussian elimination~\cite{vonManteuffel:2019wbj}.
We then reconstructed new template equations using \texttt{FiniteFlow}’s sparse solver and, from them, generated a substantially smaller system of equations. This final system of equations is fed into our finite-field workflow, where we solve it on the fly using the traditional Laporta algorithm~\cite{Laporta:2001dd} to reduce the amplitudes onto MIs.

\smallskip 

Finally, reducing the number of sample points required for the reconstruction is where we can leverage the most our physical understanding of the analytic structure of scattering amplitudes. This number is strictly related to the complexity of the rational coefficients to be reconstructed: the higher are the numerator/denominator polynomial degrees of the coefficients, the more points are necessary. We must therefore reduce the complexity of the coefficients which need to be reconstructed. We employ a multi-step strategy, discussed thoroughly in ref.~\cite{Badger:2021imn}.

\textbf{Step 1}: We fit the linear relations among the $\{r_i(\vec{x})\}$, i.e.\ we find the $c_i \in \mathbb{Q}$ such that
\begin{align} \label{eq:linrels}
\sum_i c_i  \, r_i(\vec{x}) = 0 \quad \forall \ \vec{x} \,.
\end{align}
The fit requires a number of evaluations of the coefficients equal to the number of linearly independent coefficients plus a few extra evaluations for checks. This is typically considerably smaller than the number of evaluations required to reconstruct the rational coefficients. 
We then solve the obtained linear relations so as to express the more complicated coefficients in terms of simpler ones. Only the latter then need to be reconstructed.

\textbf{Step 2}: The denominators of the coefficients are intimately related to the singularities of the amplitudes and of the Feynman integrals, which are well understood. 
We can thus make the ansatz
\begin{align} \label{eq:ansatz_factors}
r_i(\vec{x}) = \frac{\mathrm{num}_i(\vec{x})}{ \prod_j W_j(\vec{x})^{q_{ij}} } \,, \qquad \quad q_{ij} \in \mathbb{Z} \,,
\end{align}
where $\mathrm{num}_i(\vec{x})$ is a polynomial, and the possible factors $W_j(\vec{x})$ are known from the computation of the Feynman integrals (the letters of the pentagon alphabet~\cite{Chicherin:2017dob}).\footnote{Spinor factors ($\langle ij \rangle$, $[ij]$) also need to be included in the ansatz to compute objects with non-trivial helicity scaling.} We determine the integers $q_{ij}$ by reconstructing the coefficients analytically on a random univariate phase-space slice~\cite{Abreu:2018zmy}, 
which is much cheaper than the full multivariate reconstruction. From the reconstruction on the univariate slice we can also determine the degrees of the numerators, which remain to be reconstructed.

\textbf{Step 3}: We perform a univariate partial fraction decomposition of the coefficients $r_i(\vec{x})$ with respect to one of the variables on the fly. In other words, we make an ansatz for the decomposition, and reconstruct the coefficients of the ansatz by solving a fit. The variable for the univariate partial fraction decomposition is chosen so as to minimise the degrees of the coefficients of the ansatz, which depend on one less variable. For example, consider the following rational function:
\begin{align} \label{eq:ex1}
r(x,y) = \frac{- 2 x^4 - 4 x^3 y + 5 x^2 y^2 - x y^3 + 4 y^4}{(x-y) y^2 (x^2+y^2)} \,.
\end{align}
Its partial fraction decomposition with respect to $y$ is given by
\begin{align} \label{eq:ex2}
r(x,y) = - \frac{2 x}{y^2} - \frac{6}{y} + \frac{1}{x-y} + \frac{3 y }{x^2 + y^2} \,.
\end{align}
We then make the following ansatz for the partial fraction decomposition of $r(x,y)$,
\begin{align} \label{eq:ex3}
r(x,y) = \frac{q_1(x)}{y^2} + \frac{q_2(x)}{y} + \frac{q_3(x)}{x-y} + \frac{q_4(x) + q_5(x) y}{x^2+y^2} \,,
\end{align}
and reconstruct the coefficients $q_i(x)$ from the numerical evaluations of the original function $r(x,y)$.
By comparing eqs.~\eqref{eq:ex2} and~\eqref{eq:ex3} we see that
\begin{align}
q_1(x) = - 2 x\,, \qquad q_2(x) = -6 \,, \qquad q_3(x) = 1 \,, \qquad q_4(x) = 0 \,, \qquad q_5(x) = 3 \,.
\end{align}
As a result, the coefficients of the ansatz have maximum numerator|denominator degrees 1|0 in one variable. This is to be contrasted with the degrees 4|5 in two variables of the expression with a common denominator in eq.~\eqref{eq:ex1}. Constructing the ansatz in eq.~\eqref{eq:ex3} requires the knowledge of the denominator and of the degree in $y$ of the numerator of $r(x,y)$. The former is known from the previous step, and the latter can be determined by reconstructing $r(x,y)$ on a univariate slice where we vary $y$ and keep $x$ constant.

\textbf{Step 4}: We carry out another factor-matching (step 2), this time on the coefficients of the ansätze for the partial fraction decompositions (the $q_i(x)$'s in eq.~\eqref{eq:ex3} for the toy example above). We enlarge the list of factors $W_i(\vec{x})$ in the ansatz~\eqref{eq:ansatz_factors} to include also the spurious factors introduced by the univariate partial fraction decomposition in the previous step.

Finally, we multiply away all factors which have been determined, and reconstruct what remains using \texttt{FiniteFlow}'s built-in functional reconstruction algorithms. In the most complicated cases, the evaluation of the coefficients in just one prime field is not sufficient to lift all constants from the finite field to the field of rational numbers. In such cases, it is necessary to combine the reconstructions over multiple prime fields using the Chinese remainder theorem. A thorough discussion of finite field methods in the context of scattering amplitudes can be found in ref.~\cite{Peraro:2019svx}.

\smallskip

I conclude by showcasing the effectiveness of this strategy in the recent computation of the two-loop helicity amplitudes required for the double-virtual NNLO QCD corrections to isolated photon production in association with two jets at the LHC~\cite{Badger:2023mgf}. The most complicated partonic channel is the 2-quark-2-gluon channel ($0\to \bar{q} q gg \gamma$). The amplitudes for this process depend on 5 independent variables. We set one to $1$ throughout the computation, and recover it at the end from dimensional analysis. We decompose each helicity amplitude into various partial amplitudes based on the colour dependence, the number of closed fermion loops, and what the photon couples to. The reduction of the polynomial degrees at each step of the reconstruction is shown in \cref{tab:deg} for the most complicated partial amplitudes. As can be seen, the maximum polynomial degree goes down from 94 in 4 variables to 22 in 3 variables thanks to our reconstruction strategy, leading to a substantial drop in the number of sample points required to complete the reconstruction.

\renewcommand{\arraystretch}{1.5}
\begin{table}
    \centering
    \begin{tabular}{cc|ccc|cc}
        & & \multicolumn{3}{c|}{4 variables} & \multicolumn{2}{c}{3 variables} \\
        \hline
        amplitude & helicity & common denominator & step 1 & step 2 & step 3 & step 4 \\
        \hline
        $A^{(2),1}_{34;q}$ & $-++-+$ & 94|91 & 74|71 & 74|0 & 22|18 & 22|0 \\
        $A^{(2),1}_{34;q}$ & $-+-++$ & 93|89 & 90|86 & 90|0 & 24|14 & 18|0 \\
        \hline
        $A^{(2),N_c^2}_{34;q}$ & $-+-++$ & 58|55 & 54|51 & 53|0 & 20|16 & 20|0 \\
        \hline
    \end{tabular}
    \caption{
        Highest numerator|denominator degrees of the rational coefficients of the two most complicated $0 \to \bar{q} q gg \gamma$ two-loop partial amplitudes $A^{(2)}$. The superscripts denote the relevant power of $N_c$, while the subscripts give further information about the partial amplitude (see ref.~\cite{Badger:2023mgf}). The last row shows the most complicated leading-colour partial amplitude for comparison.
    }
    \label{tab:deg}
\end{table}

\section{Conclusions}
\label{sec:Conclusions}

In this talk I presented an approach to compute scattering amplitudes analytically which replaces the symbolic manipulations of the intermediate expressions with numerical evaluations in a finite field, this way sidestepping the plague of intermediate expression swell. In particular, I discussed a number of strategies which leverage our understanding of the analytic structure of scattering amplitudes to optimise the functional reconstruction of the result from the numerical samples in the finite field. This approach is well suited for the computation of high-multiplicity amplitudes. Indeed, finite field methods such as the ones I presented underly the recent progress in the computation of two-loop five-particle scattering amplitudes in QCD. The computation of the latter is one of the main bottleneck towards obtaining NNLO QCD predictions for $2 \to 3$ processes, which are in high demand for the LHC programme.

As an example, I showcased the effectiveness of our approach in the computation of the full-colour two-loop five-particle amplitudes required for the NNLO QCD corrections to isolated photon production in association with two jets at the LHC~\cite{Badger:2023mgf}.
This work marked two important milestones.
First, it completed the study at NNLO in QCD of all $2\to 3$ processes of interest for the LHC whose amplitudes involve massless particles only. Second, it was the first calculation of a $2 \to 3$ hadron-collider process at NNLO in QCD that did not rely on the leading-colour approximation. 
I refer to Michał Czakon's talk~\cite{talk-czakon} for a discussion of the phenomenology of this process.

\acknowledgments

I thank Simon Badger, Michał Czakon, Heribertus Bayu Hartanto, Ryan Moodie, Tiziano Peraro and Rene Poncelet for the great work done together in ref.~\cite{Badger:2023mgf}.
This project received funding from the European Research Council (ERC) under the European Union’s Horizon 2020 research and innovation programme \textit{High precision multi-jet dynamics at the LHC} (grant agreement No.~772099), 
and from the European Union’s Horizon research and innovation programme under the Marie Skłodowska-Curie grant agreement No.~101105486.

\bibliographystyle{JHEP}
\bibliography{bibliography}

\providecommand{\href}[2]{#2}\begingroup\raggedright\begin{thebibliography}{10}

\bibitem{Abreu:2019odu}
S.~Abreu, J.~Dormans, F.~Febres~Cordero, H.~Ita, B.~Page and V.~Sotnikov,
  \emph{{Analytic Form of the Planar Two-Loop Five-Parton Scattering Amplitudes
  in QCD}}, \href{https://doi.org/10.1007/JHEP05(2019)084}{\emph{JHEP}
  {\bfseries 05} (2019) 084}
  [\href{https://arxiv.org/abs/1904.00945}{{\ttfamily 1904.00945}}].

\bibitem{Abreu:2021oya}
S.~Abreu, F.~Febres~Cordero, H.~Ita, B.~Page and V.~Sotnikov,
  \emph{{Leading-color two-loop QCD corrections for three-jet production at
  hadron colliders}},
  \href{https://doi.org/10.1007/JHEP07(2021)095}{\emph{JHEP} {\bfseries 07}
  (2021) 095} [\href{https://arxiv.org/abs/2102.13609}{{\ttfamily
  2102.13609}}].

\bibitem{Chawdhry:2019bji}
H.A.~Chawdhry, M.L.~Czakon, A.~Mitov and R.~Poncelet, \emph{{NNLO QCD
  corrections to three-photon production at the LHC}},
  \href{https://doi.org/10.1007/JHEP02(2020)057}{\emph{JHEP} {\bfseries 02}
  (2020) 057} [\href{https://arxiv.org/abs/1911.00479}{{\ttfamily
  1911.00479}}].

\bibitem{Abreu:2020cwb}
S.~Abreu, B.~Page, E.~Pascual and V.~Sotnikov, \emph{{Leading-Color Two-Loop
  QCD Corrections for Three-Photon Production at Hadron Colliders}},
  \href{https://doi.org/10.1007/JHEP01(2021)078}{\emph{JHEP} {\bfseries 01}
  (2021) 078} [\href{https://arxiv.org/abs/2010.15834}{{\ttfamily
  2010.15834}}].

\bibitem{Chawdhry:2020for}
H.A.~Chawdhry, M.~Czakon, A.~Mitov and R.~Poncelet, \emph{{Two-loop
  leading-color helicity amplitudes for three-photon production at the LHC}},
  \href{https://doi.org/10.1007/JHEP06(2021)150}{\emph{JHEP} {\bfseries 06}
  (2021) 150} [\href{https://arxiv.org/abs/2012.13553}{{\ttfamily
  2012.13553}}].

\bibitem{Abreu:2023bdp}
S.~Abreu, G.~De~Laurentis, H.~Ita, M.~Klinkert, B.~Page and V.~Sotnikov,
  \emph{{Two-Loop QCD Corrections for Three-Photon Production at Hadron
  Colliders}},  \href{https://arxiv.org/abs/2305.17056}{{\ttfamily
  2305.17056}}.

\bibitem{Agarwal:2021grm}
B.~Agarwal, F.~Buccioni, A.~von Manteuffel and L.~Tancredi, \emph{{Two-loop
  leading colour QCD corrections to $q \bar{q} \to \gamma \gamma g$ and $q g
  \to \gamma \gamma q$}},
  \href{https://doi.org/10.1007/JHEP04(2021)201}{\emph{JHEP} {\bfseries 04}
  (2021) 201} [\href{https://arxiv.org/abs/2102.01820}{{\ttfamily
  2102.01820}}].

\bibitem{Chawdhry:2021mkw}
H.A.~Chawdhry, M.~Czakon, A.~Mitov and R.~Poncelet, \emph{{Two-loop
  leading-colour QCD helicity amplitudes for two-photon plus jet production at
  the LHC}}, \href{https://doi.org/10.1007/JHEP07(2021)164}{\emph{JHEP}
  {\bfseries 07} (2021) 164}
  [\href{https://arxiv.org/abs/2103.04319}{{\ttfamily 2103.04319}}].

\bibitem{Badger:2023mgf}
S.~Badger, M.~Czakon, H.B.~Hartanto, R.~Moodie, T.~Peraro, R.~Poncelet et~al.,
  \emph{{Isolated photon production in association with a jet pair through
  next-to-next-to-leading order in QCD}},
  \href{https://arxiv.org/abs/2304.06682}{{\ttfamily 2304.06682}}.

\bibitem{Hartanto:2019uvl}
H.B.~Hartanto, S.~Badger, C.~Br\o{}nnum-Hansen and T.~Peraro, \emph{{A
  numerical evaluation of planar two-loop helicity amplitudes for a W-boson
  plus four partons}},
  \href{https://doi.org/10.1007/JHEP09(2019)119}{\emph{JHEP} {\bfseries 09}
  (2019) 119} [\href{https://arxiv.org/abs/1906.11862}{{\ttfamily
  1906.11862}}].

\bibitem{Abreu:2021asb}
S.~Abreu, F.~Febres~Cordero, H.~Ita, M.~Klinkert, B.~Page and V.~Sotnikov,
  \emph{{Leading-color two-loop amplitudes for four partons and a W boson in
  QCD}}, \href{https://doi.org/10.1007/JHEP04(2022)042}{\emph{JHEP} {\bfseries
  04} (2022) 042} [\href{https://arxiv.org/abs/2110.07541}{{\ttfamily
  2110.07541}}].

\bibitem{Badger:2021nhg}
S.~Badger, H.B.~Hartanto and S.~Zoia, \emph{{Two-Loop QCD Corrections to
  Wbb\textasciimacron{} Production at Hadron Colliders}},
  \href{https://doi.org/10.1103/PhysRevLett.127.012001}{\emph{Phys. Rev. Lett.}
  {\bfseries 127} (2021) 012001}
  [\href{https://arxiv.org/abs/2102.02516}{{\ttfamily 2102.02516}}].

\bibitem{Badger:2021ega}
S.~Badger, H.B.~Hartanto, J.~Kry\'s and S.~Zoia, \emph{{Two-loop leading-colour
  QCD helicity amplitudes for Higgs boson production in association with a
  bottom-quark pair at the LHC}},
  \href{https://doi.org/10.1007/JHEP11(2021)012}{\emph{JHEP} {\bfseries 11}
  (2021) 012} [\href{https://arxiv.org/abs/2107.14733}{{\ttfamily
  2107.14733}}].

\bibitem{Badger:2022ncb}
S.~Badger, H.B.~Hartanto, J.~Kry\'s and S.~Zoia, \emph{{Two-loop leading colour
  helicity amplitudes for W$^{±}$\ensuremath{\gamma} + j production at the
  LHC}}, \href{https://doi.org/10.1007/JHEP05(2022)035}{\emph{JHEP} {\bfseries
  05} (2022) 035} [\href{https://arxiv.org/abs/2201.04075}{{\ttfamily
  2201.04075}}].

\bibitem{Hartanto:2022qhh}
H.B.~Hartanto, R.~Poncelet, A.~Popescu and S.~Zoia,
  \emph{{Next-to-next-to-leading order QCD corrections to Wbb\textasciimacron{}
  production at the LHC}},
  \href{https://doi.org/10.1103/PhysRevD.106.074016}{\emph{Phys. Rev. D}
  {\bfseries 106} (2022) 074016}
  [\href{https://arxiv.org/abs/2205.01687}{{\ttfamily 2205.01687}}].

\bibitem{Badger:2022mrb}
S.~Badger, M.~Becchetti, E.~Chaubey, R.~Marzucca and F.~Sarandrea,
  \emph{{One-loop QCD helicity amplitudes for pp \textrightarrow{} $
  t\overline{t}j $ to O(\ensuremath{\varepsilon}$^{2}$)}},
  \href{https://doi.org/10.1007/JHEP06(2022)066}{\emph{JHEP} {\bfseries 06}
  (2022) 066} [\href{https://arxiv.org/abs/2201.12188}{{\ttfamily
  2201.12188}}].

\bibitem{Badger:2022hno}
S.~Badger, M.~Becchetti, E.~Chaubey and R.~Marzucca, \emph{{Two-loop master
  integrals for a planar topology contributing to pp \textrightarrow{}$
  t\overline{t}j $}},
  \href{https://doi.org/10.1007/JHEP01(2023)156}{\emph{JHEP} {\bfseries 01}
  (2023) 156} [\href{https://arxiv.org/abs/2210.17477}{{\ttfamily
  2210.17477}}].

\bibitem{talk-becchetti}
M.~Becchetti,
  \emph{\href{https://indico.ph.ed.ac.uk/event/118/contributions/2360/}{Two-loop
  Feynman integrals for top-quark pair plus jet production}}, {\emph{RADCOR
  2023 -- 16th International Symposium on Radiative Corrections: Applications
  of Quantum Field Theory to Phenomenology} (2023) }.

\bibitem{Kallweit:2020gcp}
S.~Kallweit, V.~Sotnikov and M.~Wiesemann, \emph{{Triphoton production at
  hadron colliders in NNLO QCD}},
  \href{https://doi.org/10.1016/j.physletb.2020.136013}{\emph{Phys. Lett. B}
  {\bfseries 812} (2021) 136013}
  [\href{https://arxiv.org/abs/2010.04681}{{\ttfamily 2010.04681}}].

\bibitem{Chawdhry:2021hkp}
H.A.~Chawdhry, M.~Czakon, A.~Mitov and R.~Poncelet, \emph{{NNLO QCD corrections
  to diphoton production with an additional jet at the LHC}},
  \href{https://doi.org/10.1007/JHEP09(2021)093}{\emph{JHEP} {\bfseries 09}
  (2021) 093} [\href{https://arxiv.org/abs/2105.06940}{{\ttfamily
  2105.06940}}].

\bibitem{Czakon:2021mjy}
M.~Czakon, A.~Mitov and R.~Poncelet, \emph{{Next-to-Next-to-Leading Order Study
  of Three-Jet Production at the LHC}},
  \href{https://doi.org/10.1103/PhysRevLett.127.152001}{\emph{Phys. Rev. Lett.}
  {\bfseries 127} (2021) 152001}
  [\href{https://arxiv.org/abs/2106.05331}{{\ttfamily 2106.05331}}].

\bibitem{Badger:2021ohm}
S.~Badger, T.~Gehrmann, M.~Marcoli and R.~Moodie, \emph{{Next-to-leading order
  QCD corrections to diphoton-plus-jet production through gluon fusion at the
  LHC}}, \href{https://doi.org/10.1016/j.physletb.2021.136802}{\emph{Phys.
  Lett. B} {\bfseries 824} (2022) 136802}
  [\href{https://arxiv.org/abs/2109.12003}{{\ttfamily 2109.12003}}].

\bibitem{Chen:2022ktf}
X.~Chen, T.~Gehrmann, E.W.N.~Glover, A.~Huss and M.~Marcoli, \emph{{Automation
  of antenna subtraction in colour space: gluonic processes}},
  \href{https://doi.org/10.1007/JHEP10(2022)099}{\emph{JHEP} {\bfseries 10}
  (2022) 099} [\href{https://arxiv.org/abs/2203.13531}{{\ttfamily
  2203.13531}}].

\bibitem{Hartanto:2022ypo}
H.B.~Hartanto, R.~Poncelet, A.~Popescu and S.~Zoia, \emph{{Flavour
  anti-$k_\text{T}$ algorithm applied to $Wb\bar{b}$ production at the LHC}},
  \href{https://arxiv.org/abs/2209.03280}{{\ttfamily 2209.03280}}.

\bibitem{Buonocore:2022pqq}
L.~Buonocore, S.~Devoto, S.~Kallweit, J.~Mazzitelli, L.~Rottoli and C.~Savoini,
  \emph{{Associated production of a W boson and massive bottom quarks at
  next-to-next-to-leading order in QCD}},
  \href{https://doi.org/10.1103/PhysRevD.107.074032}{\emph{Phys. Rev. D}
  {\bfseries 107} (2023) 074032}
  [\href{https://arxiv.org/abs/2212.04954}{{\ttfamily 2212.04954}}].

\bibitem{Alvarez:2023fhi}
M.~Alvarez, J.~Cantero, M.~Czakon, J.~Llorente, A.~Mitov and R.~Poncelet,
  \emph{{NNLO QCD corrections to event shapes at the LHC}},
  \href{https://doi.org/10.1007/JHEP03(2023)129}{\emph{JHEP} {\bfseries 03}
  (2023) 129} [\href{https://arxiv.org/abs/2301.01086}{{\ttfamily
  2301.01086}}].

\bibitem{Gehrmann:2018yef}
T.~Gehrmann, J.M.~Henn and N.A.~Lo~Presti, \emph{{Pentagon functions for
  massless planar scattering amplitudes}},
  \href{https://doi.org/10.1007/JHEP10(2018)103}{\emph{JHEP} {\bfseries 10}
  (2018) 103} [\href{https://arxiv.org/abs/1807.09812}{{\ttfamily
  1807.09812}}].

\bibitem{Chicherin:2018old}
D.~Chicherin, T.~Gehrmann, J.M.~Henn, P.~Wasser, Y.~Zhang and S.~Zoia,
  \emph{{All Master Integrals for Three-Jet Production at
  Next-to-Next-to-Leading Order}},
  \href{https://doi.org/10.1103/PhysRevLett.123.041603}{\emph{Phys. Rev. Lett.}
  {\bfseries 123} (2019) 041603}
  [\href{https://arxiv.org/abs/1812.11160}{{\ttfamily 1812.11160}}].

\bibitem{Chicherin:2020oor}
D.~Chicherin and V.~Sotnikov, \emph{{Pentagon Functions for Scattering of Five
  Massless Particles}},
  \href{https://doi.org/10.1007/JHEP12(2020)167}{\emph{JHEP} {\bfseries 20}
  (2020) 167} [\href{https://arxiv.org/abs/2009.07803}{{\ttfamily
  2009.07803}}].

\bibitem{Chicherin:2021dyp}
D.~Chicherin, V.~Sotnikov and S.~Zoia, \emph{{Pentagon functions for one-mass
  planar scattering amplitudes}},
  \href{https://doi.org/10.1007/JHEP01(2022)096}{\emph{JHEP} {\bfseries 01}
  (2022) 096} [\href{https://arxiv.org/abs/2110.10111}{{\ttfamily
  2110.10111}}].

\bibitem{Abreu:2023rco}
S.~Abreu, D.~Chicherin, H.~Ita, B.~Page, V.~Sotnikov, W.~Tschernow et~al.,
  \emph{{All Two-Loop Feynman Integrals for Five-Point One-Mass Scattering}},
  \href{https://arxiv.org/abs/2306.15431}{{\ttfamily 2306.15431}}.

\bibitem{talk-chicherin}
D.~Chicherin,
  \emph{\href{https://indico.ph.ed.ac.uk/event/118/contributions/2338/}{Pentagon
  Functions for One-Mass Scattering}}, {\emph{RADCOR 2023 -- 16th International
  Symposium on Radiative Corrections: Applications of Quantum Field Theory to
  Phenomenology} (2023) }.

\bibitem{vonManteuffel:2014ixa}
A.~von Manteuffel and R.M.~Schabinger, \emph{{A novel approach to integration
  by parts reduction}},
  \href{https://doi.org/10.1016/j.physletb.2015.03.029}{\emph{Phys. Lett. B}
  {\bfseries 744} (2015) 101}
  [\href{https://arxiv.org/abs/1406.4513}{{\ttfamily 1406.4513}}].

\bibitem{Peraro:2016wsq}
T.~Peraro, \emph{{Scattering amplitudes over finite fields and multivariate
  functional reconstruction}},
  \href{https://doi.org/10.1007/JHEP12(2016)030}{\emph{JHEP} {\bfseries 12}
  (2016) 030} [\href{https://arxiv.org/abs/1608.01902}{{\ttfamily
  1608.01902}}].

\bibitem{Badger:2017jhb}
S.~Badger, C.~Br\o{}nnum-Hansen, H.B.~Hartanto and T.~Peraro, \emph{{First look
  at two-loop five-gluon scattering in QCD}},
  \href{https://doi.org/10.1103/PhysRevLett.120.092001}{\emph{Phys. Rev. Lett.}
  {\bfseries 120} (2018) 092001}
  [\href{https://arxiv.org/abs/1712.02229}{{\ttfamily 1712.02229}}].

\bibitem{Badger:2018enw}
S.~Badger, C.~Br\o{}nnum-Hansen, H.B.~Hartanto and T.~Peraro, \emph{{Analytic
  helicity amplitudes for two-loop five-gluon scattering: the single-minus
  case}}, \href{https://doi.org/10.1007/JHEP01(2019)186}{\emph{JHEP} {\bfseries
  01} (2019) 186} [\href{https://arxiv.org/abs/1811.11699}{{\ttfamily
  1811.11699}}].

\bibitem{Badger:2019djh}
S.~Badger, D.~Chicherin, T.~Gehrmann, G.~Heinrich, J.M.~Henn, T.~Peraro et~al.,
  \emph{{Analytic form of the full two-loop five-gluon all-plus helicity
  amplitude}},
  \href{https://doi.org/10.1103/PhysRevLett.123.071601}{\emph{Phys. Rev. Lett.}
  {\bfseries 123} (2019) 071601}
  [\href{https://arxiv.org/abs/1905.03733}{{\ttfamily 1905.03733}}].

\bibitem{Peraro:2019svx}
T.~Peraro, \emph{{FiniteFlow: multivariate functional reconstruction using
  finite fields and dataflow graphs}},
  \href{https://doi.org/10.1007/JHEP07(2019)031}{\emph{JHEP} {\bfseries 07}
  (2019) 031} [\href{https://arxiv.org/abs/1905.08019}{{\ttfamily
  1905.08019}}].

\bibitem{Badger:2021owl}
S.~Badger, E.~Chaubey, H.B.~Hartanto and R.~Marzucca, \emph{{Two-loop leading
  colour QCD helicity amplitudes for top quark pair production in the gluon
  fusion channel}}, \href{https://doi.org/10.1007/JHEP06(2021)163}{\emph{JHEP}
  {\bfseries 06} (2021) 163}
  [\href{https://arxiv.org/abs/2102.13450}{{\ttfamily 2102.13450}}].

\bibitem{Badger:2023xtl}
S.~Badger, J.~Kry\'s, R.~Moodie and S.~Zoia, \emph{{Lepton-pair scattering with
  an off-shell and an on-shell photon at two loops in massless QED}},
  \href{https://arxiv.org/abs/2307.03098}{{\ttfamily 2307.03098}}.

\bibitem{Mangano:1990by}
M.L.~Mangano and S.J.~Parke, \emph{{Multiparton amplitudes in gauge theories}},
  \href{https://doi.org/10.1016/0370-1573(91)90091-Y}{\emph{Phys. Rept.}
  {\bfseries 200} (1991) 301}
  [\href{https://arxiv.org/abs/hep-th/0509223}{{\ttfamily hep-th/0509223}}].

\bibitem{Dixon:1996wi}
L.J.~Dixon, \emph{{Calculating scattering amplitudes efficiently}},  in
  \emph{{Theoretical Advanced Study Institute in Elementary Particle Physics
  (TASI 95): QCD and Beyond}}, pp.~539--584, 1, 1996
  [\href{https://arxiv.org/abs/hep-ph/9601359}{{\ttfamily hep-ph/9601359}}].

\bibitem{Badger:2023eqz}
S.~Badger, J.~Henn, J.~Plefka and S.~Zoia, \emph{{Scattering Amplitudes in
  Quantum Field Theory}},  \href{https://arxiv.org/abs/2306.05976}{{\ttfamily
  2306.05976}}.

\bibitem{Nogueira:1991ex}
P.~Nogueira, \emph{{Automatic Feynman graph generation}},
  \href{https://doi.org/10.1006/jcph.1993.1074}{\emph{J. Comput. Phys.}
  {\bfseries 105} (1993) 279}.

\bibitem{Hodges:2009hk}
A.~Hodges, \emph{{Eliminating spurious poles from gauge-theoretic amplitudes}},
  \href{https://doi.org/10.1007/JHEP05(2013)135}{\emph{JHEP} {\bfseries 05}
  (2013) 135} [\href{https://arxiv.org/abs/0905.1473}{{\ttfamily 0905.1473}}].

\bibitem{Badger:2016uuq}
S.~Badger, \emph{{Automating QCD amplitudes with on-shell methods}},
  \href{https://doi.org/10.1088/1742-6596/762/1/012057}{\emph{J. Phys. Conf.
  Ser.} {\bfseries 762} (2016) 012057}
  [\href{https://arxiv.org/abs/1605.02172}{{\ttfamily 1605.02172}}].

\bibitem{Tkachov:1981wb}
F.V.~Tkachov, \emph{{A Theorem on Analytical Calculability of Four Loop
  Renormalization Group Functions}},
  \href{https://doi.org/10.1016/0370-2693(81)90288-4}{\emph{Phys. Lett. B}
  {\bfseries 100} (1981) 65}.

\bibitem{Chetyrkin:1981qh}
K.G.~Chetyrkin and F.V.~Tkachov, \emph{{Integration by Parts: The Algorithm to
  Calculate beta Functions in 4 Loops}},
  \href{https://doi.org/10.1016/0550-3213(81)90199-1}{\emph{Nucl. Phys. B}
  {\bfseries 192} (1981) 159}.

\bibitem{Laporta:2001dd}
S.~Laporta, \emph{{High precision calculation of multiloop Feynman integrals by
  difference equations}},
  \href{https://doi.org/10.1142/S0217751X00002159}{\emph{Int. J. Mod. Phys. A}
  {\bfseries 15} (2000) 5087}
  [\href{https://arxiv.org/abs/hep-ph/0102033}{{\ttfamily hep-ph/0102033}}].

\bibitem{Henn:2013pwa}
J.M.~Henn, \emph{{Multiloop integrals in dimensional regularization made
  simple}}, \href{https://doi.org/10.1103/PhysRevLett.110.251601}{\emph{Phys.
  Rev. Lett.} {\bfseries 110} (2013) 251601}
  [\href{https://arxiv.org/abs/1304.1806}{{\ttfamily 1304.1806}}].

\bibitem{Henn:2014qga}
J.M.~Henn, \emph{{Lectures on differential equations for Feynman integrals}},
  \href{https://doi.org/10.1088/1751-8113/48/15/153001}{\emph{J. Phys. A}
  {\bfseries 48} (2015) 153001}
  [\href{https://arxiv.org/abs/1412.2296}{{\ttfamily 1412.2296}}].

\bibitem{Gluza:2010ws}
J.~Gluza, K.~Kajda and D.A.~Kosower, \emph{{Towards a Basis for Planar Two-Loop
  Integrals}}, \href{https://doi.org/10.1103/PhysRevD.83.045012}{\emph{Phys.
  Rev. D} {\bfseries 83} (2011) 045012}
  [\href{https://arxiv.org/abs/1009.0472}{{\ttfamily 1009.0472}}].

\bibitem{Ita:2015tya}
H.~Ita, \emph{{Two-loop Integrand Decomposition into Master Integrals and
  Surface Terms}},
  \href{https://doi.org/10.1103/PhysRevD.94.116015}{\emph{Phys. Rev. D}
  {\bfseries 94} (2016) 116015}
  [\href{https://arxiv.org/abs/1510.05626}{{\ttfamily 1510.05626}}].

\bibitem{Larsen:2015ped}
K.J.~Larsen and Y.~Zhang, \emph{{Integration-by-parts reductions from unitarity
  cuts and algebraic geometry}},
  \href{https://doi.org/10.1103/PhysRevD.93.041701}{\emph{Phys. Rev. D}
  {\bfseries 93} (2016) 041701}
  [\href{https://arxiv.org/abs/1511.01071}{{\ttfamily 1511.01071}}].

\bibitem{Baikov:1996rk}
P.A.~Baikov, \emph{{Explicit solutions of the three loop vacuum integral
  recurrence relations}},
  \href{https://doi.org/10.1016/0370-2693(96)00835-0}{\emph{Phys. Lett. B}
  {\bfseries 385} (1996) 404}
  [\href{https://arxiv.org/abs/hep-ph/9603267}{{\ttfamily hep-ph/9603267}}].

\bibitem{Baikov:1996cd}
P.A.~Baikov, \emph{{Explicit solutions of n loop vacuum integral recurrence
  relations}},  \href{https://arxiv.org/abs/hep-ph/9604254}{{\ttfamily
  hep-ph/9604254}}.

\bibitem{Bohm:2017qme}
J.~B\"ohm, A.~Georgoudis, K.J.~Larsen, M.~Schulze and Y.~Zhang, \emph{{Complete
  sets of logarithmic vector fields for integration-by-parts identities of
  Feynman integrals}},
  \href{https://doi.org/10.1103/PhysRevD.98.025023}{\emph{Phys. Rev. D}
  {\bfseries 98} (2018) 025023}
  [\href{https://arxiv.org/abs/1712.09737}{{\ttfamily 1712.09737}}].

\bibitem{vonManteuffel:2019wbj}
A.~von Manteuffel and R.M.~Schabinger, \emph{{Quark and gluon form factors in
  four loop QCD: The $N_f^2$ and $N_{q\gamma} N_f$ contributions}},
  \href{https://doi.org/10.1103/PhysRevD.99.094014}{\emph{Phys. Rev. D}
  {\bfseries 99} (2019) 094014}
  [\href{https://arxiv.org/abs/1902.08208}{{\ttfamily 1902.08208}}].

\bibitem{Badger:2021imn}
S.~Badger, C.~Br\o{}nnum-Hansen, D.~Chicherin, T.~Gehrmann, H.B.~Hartanto,
  J.~Henn et~al., \emph{{Virtual QCD corrections to gluon-initiated diphoton
  plus jet production at hadron colliders}},
  \href{https://doi.org/10.1007/JHEP11(2021)083}{\emph{JHEP} {\bfseries 11}
  (2021) 083} [\href{https://arxiv.org/abs/2106.08664}{{\ttfamily
  2106.08664}}].

\bibitem{Chicherin:2017dob}
D.~Chicherin, J.~Henn and V.~Mitev, \emph{{Bootstrapping pentagon functions}},
  \href{https://doi.org/10.1007/JHEP05(2018)164}{\emph{JHEP} {\bfseries 05}
  (2018) 164} [\href{https://arxiv.org/abs/1712.09610}{{\ttfamily
  1712.09610}}].

\bibitem{Abreu:2018zmy}
S.~Abreu, J.~Dormans, F.~Febres~Cordero, H.~Ita and B.~Page, \emph{{Analytic
  Form of Planar Two-Loop Five-Gluon Scattering Amplitudes in QCD}},
  \href{https://doi.org/10.1103/PhysRevLett.122.082002}{\emph{Phys. Rev. Lett.}
  {\bfseries 122} (2019) 082002}
  [\href{https://arxiv.org/abs/1812.04586}{{\ttfamily 1812.04586}}].

\bibitem{talk-czakon}
M.~Czakon,
  \emph{\href{https://indico.ph.ed.ac.uk/event/118/contributions/2373/}{Jet and
  Photon Physics at high perturbative orders in QCD}}, {\emph{RADCOR 2023 --
  16th International Symposium on Radiative Corrections: Applications of
  Quantum Field Theory to Phenomenology} (2023) }.

\end{thebibliography}\endgroup

\end{document}